# Photochemistry of interstellar ice forming Complex Organic Molecules

Guillermo M. Muñoz Caro[1 †], Héctor Carrascosa de Lucas[1] and Rafael Martín-Doménech[1]
[1] Centro de Astrobiología, CSIC-INTA, Torrejón de Ardoz, Madrid, Spain
[†] e-mail: munozcg@cab.inta-csic.es

**Abstract |** Astrochemistry is a well-established multidisciplinary field devoted to study molecules in space. While most astrochemists are oriented to observe molecules in the gas phase and reproduce their abundances by modeling the physical conditions of the medium, the microscopic dust particles wandering in the interstellar medium deserve the attention of a smaller community. Radiation and thermally-driven processes taking place in the bare dust, and particularly in dust particles covered by ice mantles, are mimicked in the laboratory. In addition to water, interstellar ice contains other simple species. In this Review we present our current knowledge on ice photochemistry and thermal processing that ultimately leads to formation of complex organic molecules (COMs). Numerous COMs are of astrobiological interest and match those present in comets and asteroids. Upon impact of these minor bodies, water and COMs were delivered to the earth and might have intervened in the first prebiotic reactions.

**Introduction**

The broad space regions that harbor gas and dust between stars conform the interstellar medium (ISM). The density of interstellar matter varies from a highly diluted plasma to cold dense regions. Most of the ISM is hostile environment for molecule formation due to the low densities and temperatures, and the preservation of any formed species is limited by radiation. The particle density in diffuse interstellar clouds, up to $10^3$ cm$^{-3}$, and the ultraviolet (UV) field of $10^7$ photons cm$^{-2}$ s$^{-1}$ destroys molecules in the gas phase[1]. But the interiors of dense clouds, with densities up to $10^6$ cm$^{-3}$, are protected from external UV radiation and reach temperatures near 10 K, allowing the formation of ice mantles on top of dust grains, while some degree of ice processing is still expected by direct impact of cosmic-rays and mostly by secondary-UV photons[2]. The UV-field of no more than $10^4$ photons cm$^{-2}$ s$^{-1}$ in dense clouds is generated by cosmic-ray excitation of $H_2$ molecules[3,4]. Molecular hydrogen, by far the most abundant molecule in space, can react in the gas phase to produce the simple species detected in interstellar clouds. But the presence of larger molecules cannot be accounted for with gas-phase models, and therefore solid-state chemistry operating in the cold dust must be considered as well. In dense clouds where ice-covered dust is submitted to radiation, the formation of molecules can thrive. Young stars born in dense clouds are often surrounded by disks made of gas and dust that will eventualy evolve toward planetary systems. Planets like earth were seeded with water and organic matter upon impact with comets, asteroids and small particles. It is therefore believed that prebiotic chemistry benefited from such extraterrestrial delivery of molecules produced in previous stages[5,6,7,8].

The abundant ice components $H_2O$, $CH_4$, $CO_2$, and $NH_3$ constitute the ¨polar ice phase¨, and form mainly in cold grains where atomic H- and O-rich gas accretes, while CO molecules can be formed efficiently in the gas phase and freeze out on these grains when the temperature is low enough. After the CO freeze out, $H_2CO$, $CH_3OH$, OCN$^-$, and maybe HNCO and more $CO_2$ are formed in the so-called ¨apolar ice¨ phase[9]. In addition, interstellar dust grains act as microscopic chemical reactors expected to account for a large fraction of the complex organic molecules (COMs, species with at least 6 atoms, one of them being carbon)[10] detected in the gas phase. Molecules come in close contact and react at the bare/icy dust surface[11,12], where the excess energy of the reaction is absorbed by the dust particle. This activation energy, when present, is overcome by radiation energy, or by dust heating in the warmer regions like hot cores/corinos[13-16,10]. The interaction of the bare mineral surface with the reacting species can boost the chemical complexity of the end products[17,18].

This Review presents and discusses the chemical processes in the ice exposed to vacuum-ultraviolet radiation. Experimental simulations using ice analogs and astronomical observations go hand in hand, the later are commented at the end of this Review. Thermal desorption of the ice is introduced here when we refer to the ejection of COMs and because of the role heating plays in the formation of these species in the ice[19-22]. The short section titled ¨Other types of radiation for ice processing in space¨ mentions X-rays and ions, which are not covered in this Review. There are numerous references for ion processing of the ice to simulate direct cosmic-ray impact and the comparison with UV irradiation[23-28]. Finally, atomic beams that aim to reproduce reactions on the dust surface[29-31] are also out of the scope of this review.

The ejection of molecules from the ice to the gas phase can be driven by radiation (known as photodesorption, despite its more physical nature it is described in this Review because it is intimately connected to the ice photochemistry of the ice, albeit in Box 1 to separate it from the main body of the article). Electron-promoted desorption (abbreviated as EPD)[32] and heavy ion sputtering[27] are not discussed here.

Finally, we will refer to the *organic residue* as the material that remains in the vacuum chamber at room temperature containing the products of ice irradiation and warm-up, many of them are COMs of prebiotic interest. Figure 1 depicts the processes that ice mantles undergo in inter- and circum-stellar regions leading to formation of numerous COMs. This review coincides with the detection of COMs using ALMA and other radiotelescopes, the first astronomical ice observations performed with JWST, the ongoing identification of cometary species from ESA-Rosetta data, and the asteroid sample return mission Hayabusha2 that identified numerous organic species.

After fifty years of experimentation with astrophysical ice analogs, it is not possible to provide here all the references, other reviews and books can be consulted[33-37].

**Mimicking interstellar photochemical processes of ice in the lab**
The UV field at 100 AU from the protostar, during the T-Tauri phase, can reach $10^{12}$ UV photons $cm^{-2}$ $s^{-1}$, i.e. $10^5$ times higher than the interstellar radiation field, and the disk temperature below 60 K allows ice accretion[38]. With such a UV intensity, for a typical UV cross section of $2\times10^{-18}$ $cm^2$, a molecule in the ice mantle will absorb about 1 UV photon per week[39-41]. Experimental set-ups, where ice is grown by deposition of gas onto a cold substrate near 10 K, are conceived to simulate this processing in only a few hours by increasing the UV-flux. The UV-emission provided by a microwave-discharge $H_2$ lamp (MDHL), is of the order of $10^{14}$ UV photons $cm^{-2}$ $s^{-1}$, with main bands at Lyman-α (121.6 nm) and $H_2$ lines at 157.8 and 160.8 nm to mimic the secondary-UV field in dense cloud interiors that host icy dust grains[42]. These set-ups operate under high vacuum (HV, $10^{-7}$-$10^{-8}$ mbar) or ultra-high vacuum (UHV, $10^{-10}$-$10^{-11}$ mbar) conditions. HV allows to grow and irradiate micron-thick ice layers to increase the amount of organic residue. UHV allows the study of monolayer-thick ice, where one monolayer (ML) is about 0.32 nm or a column density of 1 10$^{15}$ molecules cm$^{-2}$. The column density is the number of IR absorbing molecules per unit area of 1 cm$^2$. After irradiation the ice is gradually warmed up to increase molecular diffusion and the reactivity of species. At room temperature, the residue that contains complex organic refractory species can be extracted from the vacuum for ex situ analysis.

The physical properties of pure or multicomponent astrophysical ice analogs with a variable degree of mixing (density, porosity, binding energy, amorphous or crystalline structure) depend on ice deposition rate, oriented or background deposition, deposition and warm-up temperature, or the heating rate. Diffusion, thermal and photon-induced desorption of ice molecules, photochemistry and other processes are closely related to these ice properties[43-45]. Table 1 provides some values of these parameters. The bond-dissociation enthalpy permits to know the photon energy required to dissociate the ice components[46]. The density is intrinsic to the ice composition, structure and porosity[47-52]. Except for water ice, which presents large variations in porosity in the laboratory, there are no large discrepancies in the density values among different authors. The H$_2$CO density in Table 1 is for the liquid state, as it has not been reported in the solid[53]. The binding energy $E_b$ can be obtained empirically from $E_b$ = 30.9 x $T_{peak}$, where $T_{peak}$ is the temperature of maximum desorption during ice warm-up[54,55,21,22,56].

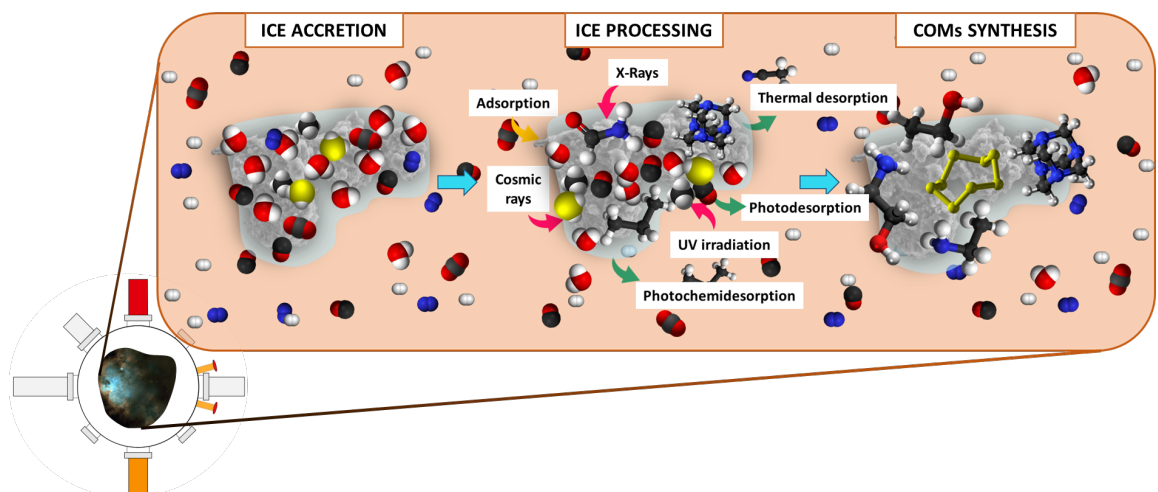

**Fig. 1. Processes expected to play a role in interstellar ice are simulated in the laboratory.** Energetic photons and ions lead to fragmentation or desorption of molecules from the ice. The end result is the synthesis of complex organic molecules (COMs) that remain in the dust, and may occasionally desorb during irradiation, or later upon warm-up as the icy dust grains experience heating in e.g., warmer regions of protoplanetary disks.

The evolution of the ice during irradiation and warm-up is monitored by infrared (IR) spectroscopy and the desorbing molecules are detected using a quadrupole mass spectrometer (QMS). The ice column density $N$, in molecules cm$^{-2}$, is obtained from integration of the IR band to deliver an absorption area $A_{int}$ and a band strength $\sigma_X$ in cm molecule$^{-1}$ specific for each vibration in molecule $X$, following $N = A_{Int}/\sigma_X$. Recently-updated band strength values are provided in Table 1[53,57,52,58,59]. Inside the vacuum chamber, IR spectroscopy monitors the formation and evolution of the residue, while gas/liquid chromatography coupled to a mass spectrometer (LC/GC-MS) and other techniques are employed for *ex situ* chemical analysis of these residues.

Figure 2 is an schematic diagram showing the formation of reactive species, radicals and ions in the irradiated ice, which further react to form new molecules higher in complexity, up to the COMs detected in the room temperature residue. Early works using GC-MS identified glycols (ethylene glycol and glycerol), carboxylic acids (being glycolic acid the most abundant one), amides (dominated by 2-hydroxycetamide and glyceramide with a smaller amount of urea), and hexamethylenetetramine[60,61]. The functional groups of these species are detected in the IR spectra of residues, showing evidence for the presence of ammonium salts of carboxylic acids $[NH_4^+][RCOO^-]$, ethers, alcohols, esters, amides, amines, and compounds related to the polyoxymethylene (POM, $(-CH_2O-)_n$) polymer. The HMT molecule detected by GC-MS is also identified by its bands in the infrared[62-64]. The decomposition of the ammonium salts of carboxylic acids inferred from IR spectra[63], upon dilution of the residues lead most likely to the carboxylic acids detected by GC-MS. Later works permitted the identification of amino acids[65-71], ribose and related sugars[72,73], and N-heterocycles besides HMT[74-78]. The nature of the residues is aliphatic, very small amounts of aromatic species were detected unless these residues are exposed to solar radiation in space during months[79,80].

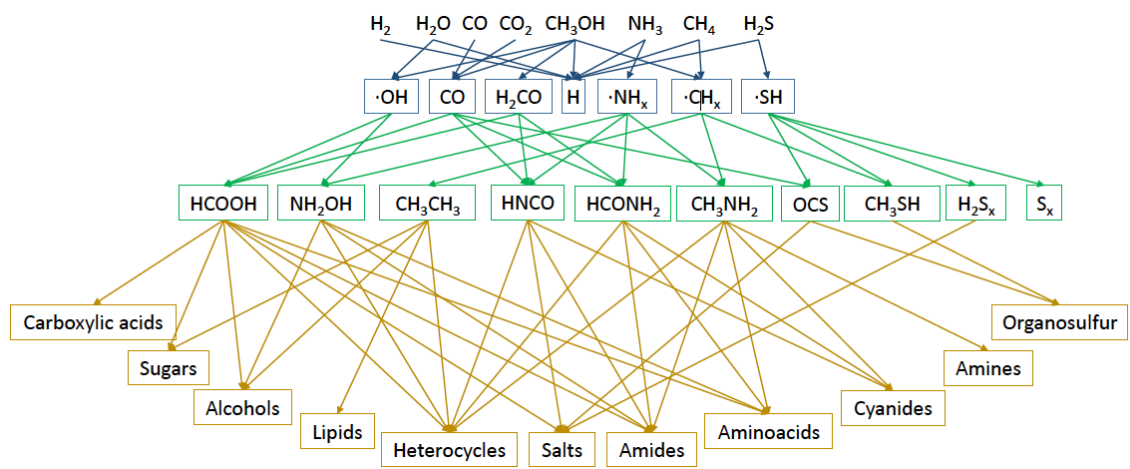

**Fig. 2. Irradiation of simple ice molecules.** These processes have been suggested among the many possible reactions that lead to photodissociation of the most abundant molecules in ice mantles shown at the top level (except $H_2S$, which was not detected in interstellar ice, but it is present in comets and is summoned as a precursor of larger S-species). This produces reactive species like $H_2CO$, ions and radicals in ice mantles (second level). Molecule-radical and radical-radical reactions produce primary species such as HCOOH, $HCONH_2$, or $CH_3SH$ (third level). Subsequent reactions can produce a full variety of prebiotic species, such as carboxylic acids, lipids, heterocycles, amino acids, etc. Arrows connect the precursors with the possible products, e.g. $HCONH_2$ may react with other species to form heterocycles, amides, aminoacids or cyanides. The final level of complexity may be reached at higher temperatures during ice warm-up. For a more detailed description of the chemical reaction network, we suggest reading the review papers cited in the Introduction and those cited in other sections for pure ices and ice mixtures, a recent compilation of the relevant reactions can be found in Muñoz Caro et al. (2019)[81].

## Box 1 | Ice photodesorption
There are several non-thermal desorption processes summoned to account for the presence of species detected in the gas of cold regions, such as cosmic rays (includes ions ranging from electrons to heavy nuclei[32,27]), spot heating[28], radical diffusion[82], or explosive chemical desorption[4]. Among them, photodesorption is the ejection of one or more molecules from the ice driven by the absorption of energetic (X-rays or UV) photons, and potentially enables their detection by radioastronomy in the gas phase. This process occurring on the ice surface will act directly on the ¨CO-rich apolar layer¨, which justifies its study in the laboratory, and prior to the formation of this apolar layer on species like $H_2O$[83-101,42,102-104]. We will see that species like $N_2$ are poor UV-absorbers but can photodesorb if mixed with CO or other molecules with a high UV-absorption cross section by energy transfer mechanisms[94,105].

The photodesorption yield of the common ice components is summarized in Table 1. Direct desorption of a molecule from the ice surface by absorption of a photon is less efficient than the desorption induced by electronic transitions (DIET). DIET occurs when the photon is absorbed by a molecule that becomes electronically excited and the photon energy is transferred to neighboring molecules, allowing the desorption of a surface molecule when intermolecular bonds are broken[87,88,90,42,106,98,105]. For CO irradiation with the MDHL, sufficient photon energy is transferred within the top 5 ML to allow photodesorption[87,88,90,42]. The photodesorption rate of solid CO measured at a specific UV monochromatic wavelength is related to the absorption at the same photon wavelength[90]. There is a rather linear decrease of the photodesorption rate with CO ice deposition temperature that might be related to dipole disorder and still needs to be better understood[88,45]. In the presence of dipole disorder, an electric field is spontaneously created in the ice. Excitation energy promotes an electron to an excited state of the molecule and leaves a "hole" behind in the ground state. The excited electron and the hole are transferred, this is known as exciton transfer. In this regard, exciton transfer in the presence of molecular dipole disorder in the ice, and the generation of a spontaneous electric field, are expected to play an important role in the photochemistry and photodesorption[107-110]. An in-depth study of CO ice photodesorption is provided in Sie et al. 2022[111].

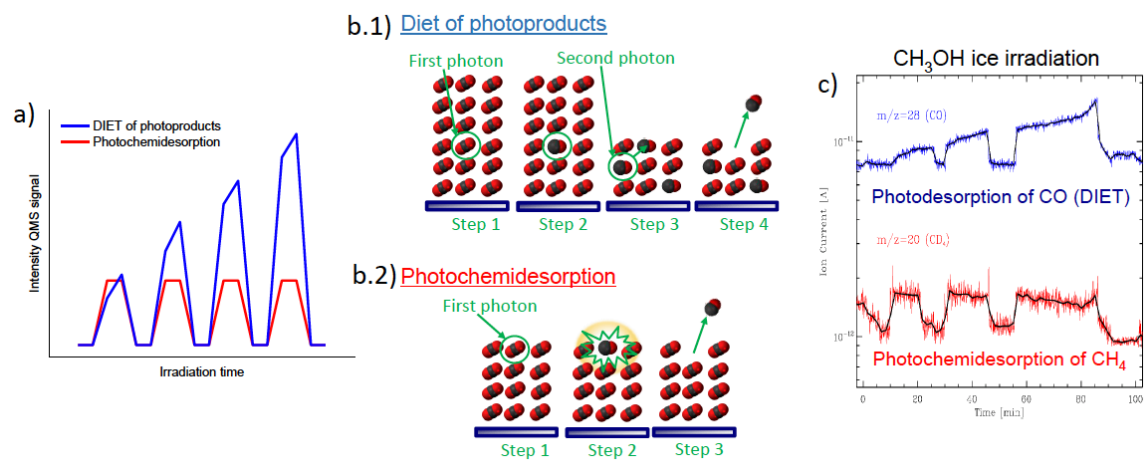

**Photodesorption processes in the ice.** a) Two patterns in the desorption of ice photoproducts were identified during ice irradiation, known as DIET of photoproducts and photochemidesorption, see Section Ice Photodesorption. b.1) The DIET of photoproducts accounts for the increasing desorption rate up to a maximum, as shown in panel a of this figure. This effect is explained by the desorption of photoproducts, whose concentration increases in the ice bulk, and become exposed at the ice surface due to photodesorption of the upper layers. Panel shows the absorption of a photon by a $CO_2$ molecule in the ice bulk, which produces a CO molecule. During continued irradiation, the desorption of the monolayers on top eventually brings this CO molecule near the ice surface, where it absorbs a new photon and energy transfer leads to the ejection of one or more CO neighbours on the surface. b.2) Photochemical desorption or photochemidesorption is related to a constant desorption rate, depicted in panel a, as the result of immediate desorption of species formed on the ice surface. Panel shows the absorption of a photon by a $CO_2$ molecule on the ice surface, forming a CO molecule that desorbs thanks to the excess energy of $CO_2$ photodissociation into CO + O. c) Experimental data collected by QMS during UV-irradiation of pure $CH_3OH$ ice is accounted for by DIET photodesorption of the CO and photochemidesorption of the $CH_4$ photoproducts, this panel is adapted from Cruz-Díaz et al. 2016[100].

Two trends were identified in the photodesorption of species made during irradiation in the ice in the laboratory, hereafter ¨photoproducts¨, an increasing photodesorption rate due to a higher concentration of photoproducts in the bulk that join the renewed ice surface, and a constant photodesorption rate during the irradiation. The later, known as photochemical desorption or photochemidesorption, permits the ejection of photoproducts formed on the ice surface supported by the excess energy of their formation[102,103,100]. This mechanism can release photoproducts from the irradiated ice that would not photodesorb following other mechanisms. For example, it works for methane molecules formed on the surface of irradiated $CH_3OH$, while photodesorption of $CH_3OH$ itself is not efficient. Attempts to photochemidesorb methanol by formation via $CH_3\bullet$ + $OH\bullet$ radicals in $H_2O:CH_4$ ice irradiation experiments were not successful, while formaldehyde did photochemidesorb in this experiment[103]. No photodesorption of COMs has been reported, but two COMs were found to photochemidesorb upon methane ice irradiation: ethane, likely formed by $CH_3\bullet$ + $CH_3\bullet$ on the ice surface, and propane by $C_2H_5\bullet$ + $CH_3\bullet$ also on the surface[112]. In summary, certain species like CO can photodesorb directly if this molecule or a nearby molecule become photoexcited, other (like $N_2$) may photodesorb aided by their neighbors, and some (like $CH_4$ and certain COMs such as ethane and propane) can photochemidesorb. The remaining species like methanol in the ice, may not be able to desorb significantly via photodesorption processes and require a different desorption mechanism. The direct cosmic-ray impact can lead to an efficient sputtering of methanol and other ice species[27,113]. Chemical desorption refers to the formation of a molecule on the ice surface that leaves the ice with no intervention of radiation in this process, some molecules like $H_2S$ can experience chemical desorption[114]. In protoplanetary disks, X-rays are expected to play a pivotal role in photochemistry and photodesorption, a dedicated experiment on this topic[115] is introduced in the last section.
Box 1 ends here

Table 1 | **Values of some physical properties of common and potential components in ice mantles.** Bond dissociation enthalpy at room temperature; these values do not change considerably at cryogenic temperatures[46]. The IR band strength (in cm molecule$^{-1}$), density (in g cm$^{-3}$) and binding energy (in K, where 1 K = 0.0083 kJ/mol) of pure ices are provided. The photodesorption yield of these ices irradiated with a MDHL are given with their references.

| Ice | Bond-dissociation enthalpy (eV/bond)$^a$ | Density (g cm$^{-3}$)$^b$ | Band strength x10$^{-17}$ (cm molecule$^{-1}$)$^c$ | Binding energy (K)$^d$ | Photodesorption yield x10$^{-3}$ (molec./ph.) | References |
|---|---|---|---|---|---|---|
| $H_2O$ | 5.15 (H-OH) | 0.6-1.1 | 20 (3280 cm$^{-1}$) | 5165 | 1.3 ± 0.2 | 101 |
| CO | 11.16 (C-O) | 0.84 | 0.87 (2139 cm$^{-1}$) | 890 | 54.0 ± 5.0 | 87 |
| $CO_2$ | 5.51 (O-CO) | 1.30 | 7.6 (2343 cm$^{-1}$) | 2605 | 0.1 ± 0.04 | 102 |
| $CH_3OH$ | 3.91 ($CH_3$-OH) 4.10 (H-$CH_2OH$) | 0.65 | 23 (3500-2750 cm$^{-1}$) | 4355 | ≤ 0.03; 0.01 | 100,95 |
| $NH_3$ | 4.51 (H-$NH_2$) | 1.1 | 2.2 (3378-3208 cm$^{-1}$) | 2965 | 2.1 (+2.1,-1.0) | 104 |
| CH$_4$ | 4.55 (H-CH$_3$) | 0.47 | 0.84 (1301 cm$^{-1}$) | 1020 | < 0.17 (MDHL); 2.3 (E > 9 eV) | 112,98 |
| H$_2$S | 3.91 (H-SH) | 0.944 | 1.69 (2547 cm$^{-1}$) | 2516 | ---- | ---- |
| H$_2$CO photochem. | 3.76 (H-CHO) | 0.81 (liquid) | 0.96 (1720 cm$^{-1}$) | 3765 | 0.04 | 103 |
| CH$_4$ photochem. | 4.55 (H-CH$_3$) | 0.47 | 0.84 (1301 cm$^{-1}$) | 1020 | 2 | 100 |
| CH$_3$CH$_3$ Photochem. | 3.81 (H-CH$_2$CH$_3$) | 0.4 | 2.42 (2972 cm$^{-1}$) | 1840 | 0.8 | 112 |
| CH$_3$CH$_2$CH$_3$ Photochem. | 4.25(H-CH$_2$CH$_2$CH$_3$), 4.15 (CH$_3$CH(-H)CH$_3$) | 0.653 | 4.46 (3000-2850 cm$^{-1}$) | 4000 | 0.24 | 112 |

[a] Dissociating bond in parenthesis[46]. [b] Refs. 47-53. [c] Note: The wavenumber position (in cm$^{-1}$) and vibrational assignment of these transitions are H$_2$O (3280, OH stretch), CO (2139, CO stretch), CO$_2$ (2343, CO stretch), CH$_3$OH (3500-2750, OH & CH stretch), NH$_3$ (3378-3208, NH stretch), CH$_4$ (1301, bending), H$_2$S (2547, HS stretch), H$_2$CO (1720, CO stretch), CH$_3$CH$_3$ (2972, CH stretch), CH$_3$CH$_2$CH$_3$ (3000-2850, CH stretch)[53,57,52,58,59]. [d] Refs. 54,55,21,22,56.

## Pure ices

The specific properties and the photochemistry of different interstellar ice analog components are presented in this section. Because in some pure ices, like CO, photodesorption can strongly compete with photochemistry, it was considered appropriate to discuss the photodesorption as well. First, the key aspects of each ice are highlighted as follows. H$_2$O: displays complex phase changes, it plays an important catalytic role on the efficient formation of COMs in ice mixtures exposed to radiation; CO: the main effect of pure CO ice irradiation is photodesorption, in mixtures the photoexcited CO molecule triggers chemical reactions; CO$_2$: irradiation produces CO and the COMs obtained when CO is mixed with H$_2$O and NH$_3$; CH$_3$OH: this efficient COMs precursor produces formaldehyde and ethylene glycol, in particular formaldehyde reacts at cryogenic temperatures providing an alternative route to radical reactions toward COMs formation; CH$_4$: during irradiation hydrogenated amorphous carbon is made, CH$_3$ radicals also form ethane and propane that photochemidesorb if produced at the ice surface (they are the largest COMs found to photodesorb), while other COMs containing O and N can be formed when CH$_4$ is mixed with H$_2$O and NH$_3$ in the ice; NH$_3$: radicals are rarely detected in IR spectra of ice during irradiation but the NH and NH$_2$ radicals are an exception, their reactivity at different temperatures can be monitored; H$_2$S and other S-species: S-chains are produced by H$_2$S ice irradiation, they could be a source of the missing S in dense clouds. This Section focuses on one-component ice analogs before the photochemistry of multicomponent ices is discussed.

## Water

Amorphous solid water (ASW) is the most common type of ice in interstellar and cometary environments[116,9,117]. Atmospheric water vapor is also the main contaminant in vacuum chambers. Only under optimal ultra-high vacuum conditions is the ice surface relatively free from background water accretion, allowing the study of surface processes in ices containing other components. Photodissociation of H$_2$O ice in the first absorption band at 130–165 nm involves mainly two primary processes[118]: H$_2$O + hv → H + OH(v=0 and 1) and H$_2$O + hv → H$_2$ + O($^1$D). Water photodissociation into H + OH is often followed by recombination back to H$_2$O[119]. The photodesorption of water ice was measured with a microbalance[83]. The translational and rotational energies of photodesorbed water molecules have been reported[120]. Photodesorption of H$_2$, O$_2$, and a fraction of the formed OH radicals during water irradiation has been observed[121,101,97]. These works estimate that the photodesorption of water molecules in the ice near 10 K is within (0.5-2) x $10^{-3}$ molecules per incident VUV photon and increases with irradiation temperature, along with the photodesorption of the observed photoproducts. The $O_2$ photodesorption increases with irradiation time, as the result of building-up and diffusion of $O_2$ molecules in water ice that eventually reach the surface, which denotes a fluence dependence in $O_2$ photodesorption[101]. There is evidence for indirect desorption of weakly adsorbed species on the surface of water ice via collisions with energetic H atoms, produced by photodissociation of water molecules[122].

**Carbon monoxide**
The relative-to-water ice abundance of CO is up to 25% in quiescent clouds[9]. It has the strongest bond among all neutral species, 11.16 eV, and is therefore not dissociated by Ly-α photons, 10.2 eV. Indirect dissociation occurs in the ice irradiated with MDHL via CO + CO* → $CO_2$ + C), which along with $C_3O$ and $C_3O_2$ formation only accounts for less than 5% of the absorbed photon energy[84,87]. This enables the efficient photodesorption of CO molecules summarized in Sect. Ice Photodesorption. X-ray photons are capable of forming the CO photoproducts $C_xO/C_xO_2$ and $C_y$ with x ≤ 7 and y ≤ 10[123]; several of these products are common to keV-electron bombardment of CO ice[124].

**Carbon dioxide**
$CO_2$ can present a relative-to-water abundance up to 28% in ice mantles[9]. Various experimental works were devoted to the properties of $CO_2$ ice[125,126]. The presence of $CO_2$ in interstellar ice mantles does not correspond to the amorphous structure in segregated $CO_2$ ice measured in the laboratory[127]. This linear molecule has an interesting IR spectroscopy that traces day cycle variations in icy moons Europa and Ganymede[128]. There is ample literature on $CO_2$ ice irradiation experiments[86,89,93,96,129]. The formation of CO is the main outcome of $CO_2$-ice UV-photoprocessing using the MDHL, followed by $CO_3$, $O_2$, and $O_3$. During UV-irradiation of $CO_2$, DIET-photodesorption of CO, $O_2$ and $CO_2$ takes place through a process called indirect desorption induced by electronic transitions (DIET), with maximum photodesorption yields of $1.2 \times 10^{-2}$, $9.3 \times 10^{-4}$, and $1.1 \times 10^{-4}$ molecules/incident photon, respectively[102]. In addition, Sie et al. (2019)[130] proposed that C atoms account for about 33% of the amount of depleted $CO_2$ molecules in the ice. These authors report a more efficient photodesorption of species for different emission spectra of the MDHL.

**Methanol**
The estimated relative-to-water abundance of methanol is no more than 9% in ice mantles[9]. Among the common ice components, methanol is undoubtedly the most efficient precursor of COMs upon irradiation of interstellar ice analogs. The relatively complex photochemistry of $CH_3OH$ ice produces several COMs, such as $CH_3CH_2OH$, $CH_3OCH_3$, HCO-bearing molecules, and its dimer ethylene glycol $(CH_2OH)_2$; this work also provides the binding energies of these species[85]. The list of methanol photoproducts was enlarged thanks to their in situ GC-MS detection[131], concluding that with the exception of small molecules (mainly CO, $CH_4$, $CO_2$, and $H_2CO$), alcohols are the most abundant photoproducts, followed by aldehydes/ketones, esters and carboxylic acids. Similar products are formed by X-ray irradiation of $CH_3OH$, where the IR ice spectrum at temperatures larger than methanol desorption is well matched with that of ethylene glycol that is likely formed by reaction of 2 $CH_2OH$ (the IR band of this radical is at 1195 $cm^{-1}$)[132,133]. The IR band of the HCO radical around 1852 $cm^{-1}$ is detected in the UV and X-ray experiments and likely reacts to form HCOOH and other species[81]. The methoxy radical $CH_3O$ absorbs near 1040 and 2820 cm$^{-1}$ and overlaps with methanol bands, thus hindering its identification in the ice[134], it might precede the formation of some photoproducts, e.g. CH$_3$O + CH$_2$OH → CH$_3$OCH$_2$OH[135]. The two-step dehydrogenation of irradiated methanol molecules in the ice to form formaldehyde is of paramount importance as this reactive neutral species triggers several COM formation pathways.

**Methane**
Methane ice, with relative-to-water abundances of 5% and 2% in low-mass and high-mass protostars, respectively[136,137,9] can be formed in various ways: hydrogenation of C atoms on the dust surface, CH$_3$OH ice irradiation, or gas phase formation followed by freeze-out on dust[138]. CH$_4$ ice is observed in different objects of the solar system such as Titan[139], Triton[140], and Pluto[141].

The methane photodissociation energy is 4.58 eV (270.5 nm) in the gas phase. UV-photochemistry of methane ice produces CH$_3$ and H, and the molecules H$_2$, C$_2$H$_2$, C$_2$H$_4$, C$_2$H$_6$, and C$_3$H$_8$[142,112]. CH$_3$ radicals were only observed at photon wavelengths of 140 nm or shorter (8.57 eV). At temperatures near 10 K, rather than the CH$_3$ + CH$_3$ reaction forming C$_2$H$_6$, it was suggested that there is a higher probability for CH$_3$ radicals to react with the more abundant CH$_4$ molecules to produce C$_2$H$_6$ + H, or recombine with H atoms[143]. However, using electron paramagnetic resonance (EPR) technique for the detection of radicals in the ice, it was found that the major loss channel of CH$_3$ radicals is due to the three-particle reaction of the radical recombination with participation of a CH$_4$ molecule:
CH$_3$ + CH$_3$ + (CH$_4$) → C$_2$H$_6$ + (CH$_4$)[142]. This reaction is the major source of C$_2$H$_6$ molecules according to these authors. Abplanalp et al. (2018)[144] report a detailed reaction scheme for methane ice exposed to electrons or VUV-irradiation. They provide evidence for the formation of internally excited ethane via methyl radicals but also by two methane molecules after hydrogen elimination: (CH$_4$)$_2$ → (C$_2$H$_6$)* + H$_2$/2H. In addition to the aforementioned photoproducts, these authors identified the C$_2$H$_5$ radical, while C$_2$H$_6$, C$_3$H$_8$, C$_4$H$_{10}$ and larger saturated/unsaturated hydrocarbons containing up to 11 carbon atoms were detected during the TPD.

Hydrogenated amorphous carbon (*a*-C:H) is produced during methane ice irradiation in the laboratory at 10 K; this material reproduces the Diffuse Interstellar Medium absorption of carbon grains observed in our Galaxy and in other galaxies[145]. The *a*-C:H composition that provided the best match of the interstellar IR band consists of aliphatic chains with CH$_2$/CH$_3$ ratio around 2, connecting aromatic units of 1-2 rings[145].

Methane photodesorption was observed for monochromatic photon energies above 9 eV[98]. While there is no evidence for the direct photodesorption of methane using continuum/non-monochromatic sources as the MDHL, during methane irradiation the photodesorption of C$_2$H$_2$, and the photochemidesorption of C$_2$H$_6$ and C$_3$H$_8$ was found. As already mentioned, photochemidesorption of methane molecules produced during methanol irradiation is possible[100].

**Ammonia**
The inferred NH$_3$ abundance in ice mantles is no more than 6% with respect to water[9]. Photolysis experiments on ammonia trapped in solid matrix have shown that reactions
NH$_3$ + hν → NH$_2$ + H,

$NH_3 + h\nu \rightarrow NH + H_2$,
$NH_3 + h\nu \rightarrow NH + 2H$
are possible at photon energies above 8 eV[146]. Loeffler & Baragiola (2010)[147] found that the first two reactions are occurring during ammonia ice irradiation using 193 nm (6.42 eV) photons, Using X-ray absorption spectroscopy and 150 eV photons, it was observed that the first reaction is preferred[148]. Furthermore, the concentration of $NH_2$ radicals decreases with the irradiation dose, whereas a large amount of $NH_3$ remains available to produce $NH_2$. Since more $NH_2$ radicals are lost than formed, this suggests that radical-radical $NH_2 + NH_2$ reactions are efficient and form hydrazine:
$NH_2 + NH_2 \rightarrow N_2H_4$.
The yield of this reaction increases when the concentration of $NH_2$ becomes high enough to allow the recombination of two neighboring $NH_2$ radicals, which only occurs at high doses, 23 eV/molecule[148]. $N_2H_4$ subsequently dehydrogenates to $N_2H_2$, that can be dissociated into $N_2$ and $H_2$ molecules. Formation of $N_2H_2$, $N_2$, and $H_2$ was also reported upon 193 nm photon irradiation of $NH_3$ ice[147].

The photochemistry and photodesorption of ammonia ice irradiated with a MDHL was investigated[104]. The IR bands of $NH_2$ and $NH$ radicals around 1500 $cm^{-1}$ and 3100 $cm^{-1}$, respectively, allowed monitoring of these radicals during irradiation and warmup. The photoproducts $H_2$, $N_2$, and $N_2H_4$ were detected in the gas phase during thermal desorption of the previously irradiated $NH_3$ ice. An increasing photodesorption yield of $H_2$ and $N_2$ with fluence was also observed, while $NH_3$ molecules photodesorbed at constant rate of [2.1-1.0,2.1+2.1] $10^{-3}$ molecules/incident photon.

**Sulfur molecules**
OCS is the only sulfur-species likely identified in interstellar ices[149], with $SO_2$ being tentatively detected[150]. $H_2S$ is expected to form by hydrogenation of atomic S impinging on dust, but the band of solid $H_2S$ at 2548 $cm^{-1}$ overlaps with one of methanol and hinders its detection in interstellar ice. Nevertheless, $H_2S$ is the most abundant S-bearing molecule in comets[151], and is used in most experiments of sulfur ice photochemistry. Due to its particularly high absorption cross section in the UV, $H_2S$ readily produces HS radicals. Subsequent radical-radical reactions at 10 K produce $H_2S_2$ efficiently[152,54,153]. This process is also observed in $H_2S$ ice exposed to X-rays[154] or ions[155]. The formation of $S_x$ and $H_2S_x$ species with x from 2 to 8 upon UV-irradiation of $H_2S$ is documented[152,156,54,153,157]; these species might account for a significant amount of the ¨missing¨ sulfur in dense interstellar clouds, since the sulfur budget is very short in these clouds with respect to its cosmic value. Their identification is based on in situ QMS during the TPD thanks to the detection of their molecular ions at their corresponding desorption temperatures, and the GC-MS analysis of the residues. An efficient chemical desorption of $H_2S$ has been reported[114].

Irradiation of $SO_2$ ice with monochromatic UV photons under 200 nm at 11 K leads to the formation of $SO_3$ along with traces of $O_3$[158]. Formation of $SO_3$ in $SO_2$ ices is also observed upon X-ray[159] and proton[160] processing. An upper limit of 0.25 molecules/photon was reported for the $SO_2$ X-ray photodesorption yield[159].

In laboratory experiments, OCS is formed upon UV photon and ion irradiation of $H_2S/SO_2:CO/CO_2$ ice mixtures[161,155,162], or in $CS_2$-bearing ices with an O donor[163,164]. Ikeda et al. (2008)[165] studied the photodissociation dynamics at 90 K of OCS and $CS_2$ adsorbed on $H_2O$-ice films using 193 nm photons. The interaction of OCS with the ice film enhanced the formation of S($^3$P) over S($^1$D) compared to gas-phase photodissociation. Reaction of the photoproduced S atoms with the parent molecules led to formation of S$_2$ as a secondary product. Formation of OCS through the CS + O pathway is more favourable than through the CO + S pathway[166].

In addition to OCS, UV or electron-driven chemistry of CS$_2$-bearing, CO$_2$-, CO-, and also H$_2$O-rich ices produces a variety of S-bearing products including SO, CS, S, and a significant fraction of undetectable sulfur allotropes[166]. Finally, irradiation with 193 nm photons of CS$_2$ in a N$_2$ or Ar ice matrix at 13 K leads to isomerization from linear to cyclic CS$_2$[167]. On the other hand, Maity & Kaiser 2013)[164] reported formation of CS$_x$ (x=3-6) and C$_x$S$_y$ (x=2-3,y=1-2) molecules upon irradiation of CS$_2$ ice at 12 K with 5 keV electrons.

**Binary ice mixtures with water**

As earlier mentioned, water is the dominant component in interstellar and cometary ice mantles. For completeness, the formation of simple photoproducts detected in binary mixtures of other ice components with water is summarized in this section, these are CO, CO$_2$, CH$_3$OH, CH$_4$, NH$_3$, and H$_2$S. H$_2$O:CO ice exposed to UV and X-ray produces CO$_2$, HCO, H$_2$CO, HCOOH, (CH$_2$OH)$_2$, and CH$_3$CHO[168,81]. In CO$_2$:H$_2$O ice, H$_2$O$_2$ is produced as in the pure water ice, while formation of the species CO, CO$_3$, and O$_3$ formed in pure CO$_2$ ice is highly suppressed[169].

For CH$_3$OH:H$_2$O the products are common to the pure methanol ice UV-irradiation, these are H$_2$CO, CO, CO$_2$, CH$_4$, and (CH$_2$OH)$_2$, but the production of CH$_3$OCHO, H$_2$CO, CO$_2$, and CO is enhanced in the ice mixture, resulting from cross reactions between the products of methanol with those of water (OH and H$_2$O$_2$)[170,171]. In the CH$_4$:H$_2$O ice, the molecules CO, H$_2$CO, CH$_3$OH, CH$_3$CHO and CH$_3$CH$_2$OH are formed[172]. H$_2$O:NH$_3$ ice irradiation leads to formation of the ammonium ion (NH$_4^+$)[173]. In H$_2$S:H$_2$O ice, the UV-photoproducts SO$_2$, H$_2$SO$_2$, and H$_2$SO$_4$ were identified[54].

**Selected ice mixtures**

After the presentation of the photochemistry of pure ice components, this section discusses the experimental results on the photoprocessing of multicomponent ice samples containing H$_2$O:CO:CO$_2$:CH$_3$OH:NH$_3$:CH$_4$, to comment on intermolecular interactions and COMs formation. Upon UV-irradiation of these ice mixtures, IR bands of the radicals HCO•, •NH$_2$, and HOCO• are detected at 10 K[174]. The formation of species in the ice for different ice compositions is considered. The water content in the ice is particularly critical for the efficient formation of most COMs. A higher dilution of CO, CO$_2$, CH$_3$OH and NH$_3$ in water inhibits the polymerization of reactive molecules such as formaldehyde into POM and leads to a more efficient formation of COMs in the residues made by ice irradiation[63]. Other reasons were proposed to account for the enhancement of COMs formation in water-dominated ices, including a reduction of back reactions reforming the starting ice components, the incorporation of the O in water to COMs or a more efficient trapping of the radicals made by irradiation in the ice[172]. The water ice matrix enhances diffusion of embedded species, increasing the reactivity during structural changes of the warmed up ice, such as crystallization and pore collapse[175,176,44]. Under ultra-high vacuum, water ice desorbs when the temperature approaches 170 K and the more refractory photoproducts that remain on the substrate will have the possibility to react and evolve toward even more complex molecules. A representative example is the formation of hexamethylenetetramine that occurs after water desorption[63,177].

**Formation of polymer-like species in the ice**

Large polyoxymethylene (POM) related species are made in the warmed up ice analogs if the formaldehyde concentration is sufficiently high and a small amount of $NH_3$ is added[178]. POM formation in the ice is traced by IR spectroscopy and occurs efficiently in irradiated $CH_3OH:NH_3$ ice due to the high $H_2CO$ concentration generated in this experiment. Adding water in this mixture reduces the POM formation efficiency but it is still observed[63]. As mentioned above, a high concentration of $CH_3$ and $CH_2$ radicals in irradiated methane ice produces hydrogenated amorphous carbon already at 10 K[145]. Because the presence of water inhibits the formation of polymer-like species and favors the synthesis of COMs, water-rich ices are appealing for the development of prebiotic chemistry in the ice. The remaining sections are devoted to this topic.

## COMs of prebiotic interest made in the ice

Despite the considerable number of publications dedicated to this topic, the reaction networks forming COMs in astrophysical ice analogs is not well understood. These chemical processes in the ice are out-of-equilibrium. Quantum chemistry calculations contribute to ellucidate the reaction networks[10,179]. A possible photolytic formation of some COMs has been suggested (Bacmann et al. 2012). The reactions driving ice chemistry are radical-radical (Zhitnikov & Dmitriev 2002[142], for the case of $CH_3$ radicals with the mediation of a $CH_4$ molecule; Parent et al. 2009[148] for $NH_2$ radicals), radical-molecule (Oba et al. 2012[181] for OH + $H_2$; Borget et al. 2017[182] for CN reaction with solid $H_2$), or molecule-molecule, where reaction energy may come from temperature[183] or photons, promoting molecules to an excited state[184], with reaction barriers on the order of zero for radical-radical reactions and up to tens of kJ mol$^{-1}$ for the other reaction types, which are typically lower than the diffusion barriers of the reactants in the ice bulk[185,44]. Ice chemistry is thus diffusion limited and, therefore, the reactions favoured at the lowest temperatures near 10 K must require little or no diffusion.

Hydrogen atoms diffuse over long distances in the ice and can react by tunneling at 10 K forming $H_2O$ and organics that include $H_2CO$ and $CH_3OH$[11], leading to hydrogenation of species. A few molecular species, e.g. $H_2CO$ and HCN, can react at low temperatures, although these reactions are enhanced if diffusion is at work in the ice. Therefore, in some cases COMs are produced by warmup of certain ice mixtures in the absence of radiation. For instance, carbamic acid ($NH_2COOH$) and ammonium carbamate [$NH_2COO^-$][$NH_4^+$] are formed by warmup of $NH_3:CO_2$ ice[186], and polyoxymethylene (POM) can be produced by polymerization of $H_2CO$[178].

Acid-base reactions are barrierless and may thus occur at 10 K. A tentatively detected acid in the ice is HCOOH[187], while $NH_3$ in the form of ammonia hydrate $H_2O.NH_3$[188], or as $NH4^+$, is a likely base in the ice[189,190]. The formation of ammonium salts of carboxylic acids [(R–COO$^-$)($NH_4^+$)] occurs spontaneously at 10 K[63]. The presence of other salts in the ice was also inferred experimentally, an example is [(OCN$^-$)($NH_4^+$)][183,191].

Reactive species at cryogenic temperatures such us $H_2CO$, HCN, or HCOOH were not firmly identified in interstellar ice mantles, although $H_2CO$ and HCOOH are likely present[9]. Thermal processing of a typical ice mixture containing the firmly identified species in interstellar ice, $H_2O:NH_3:CO:CO_2:CH_3OH:CH_4$, does not lead to the formation of COMs[63]. Irradiation of interstellar ice analogs is therefore essential to increase the chemical complexity of these ices. A clear effect of ion or photon processing is the conversion of the most abundant and stable C-bearing species in the ice (CO, $CO_2$ and $CH_3OH$) to $H_2CO$[192,193]. Radicals are also abundantly formed in the energetically processed ices and participate in radical-radical or radical-molecule reactions; it is often difficult to disentangle between both processes as the end products are common. Figure 3 is a selection of relevant processes mediated by ice irradiation, the formation of primary species and the complex species detected as refractory residue components.

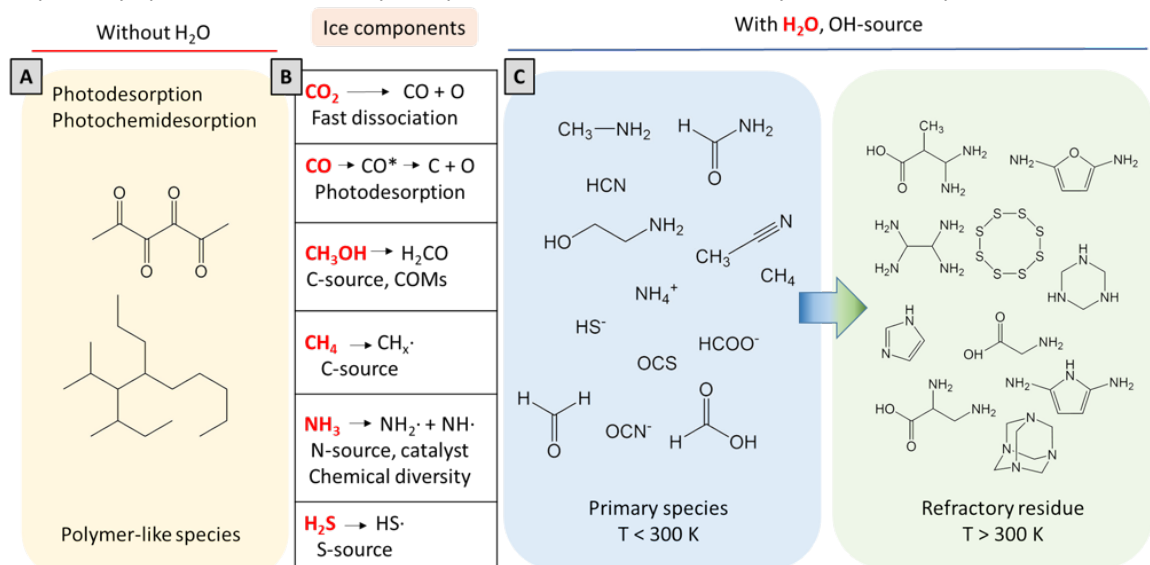

**Fig. 3. Selection of complex species detected as refractory residue components.** This figure depicts how the addition of water, the main interstellar ice component, reduces polymer formation and enables the production of numerous organic molecules. Panel B: The main processes undergone by each component of the ice mantles lead to products and determine the final composition of the residue. Photodissociation of the starting ice species produces radicals and reactive molecules like H$_2$CO. Most of the photodesorption studies were conducted in pure ice components, observing a more efficient photodesorption when the photodissociation of ice molecules is low, as in pure CO ice. Panel A: In the absence of water, photodesorption and in particular photochemidesorption of species from the ice are commonly observed. Also polymer-like materials are made in the irradiated ice at both low or high temperatures, in particular POM (from H$_2$CO) and amorphous carbon (from CH$_x$ radicals) were identified. Panel C: In water-dominated ice mixtures, the so-called ¨primary species¨ are formed at relatively low temperatures. Subsequent reactions produce the molecules that are found in the ¨refractory residue¨ at 300 K.

**Amino acids**

Two independent works were published simultaneously on the identification of racemic amino acids in residues. Bernstein et al. 2002[66] included HCN among the ice ingredients and reported the presence of glycine, serine, and alanine after hydrolysis; serine might be formed from the cycloserine detected prior to hydrolysis. Muñoz Caro et al. 2002[65] used the most common interstellar ice components (H$_2$O:CH$_3$OH:NH$_3$:CO:CO$_2$) and identified 16 amino acids. The simplest non-proteinaceous amino acid, carbamic acid (NH$_2$COOH), was also detected in the residue[193]. The relatively high abundance of diamino acids, i.e. amino acids that contain two NH$_2$ functional groups, motivated the search for these species in meteorites, and they were found in the Murchison meteorite[65,195]. Under favorable conditions, polycondensation of diamino acids could yield peptydic nucleic acid (PNA), a possible precursor of RNA and DNA[196]. The synthesis of non-racemic amino acids[197] requires a circularly polarized UV source[67,71].

The amino acid abundances in residues increase considerably upon hydrolysis, this behavior is common to meteoritic amino acids[65,66,68,198], but at least glycine[65], α-alanine, serine[68], β-alanine, and sarcosine[198] are present in the residues before hydrolysis. The potential precursors that enhance amino acid abundances after hydrolysis are discussed in the next section. The ice synthesis of amino acids in these experiments does not seem to follow a single reaction pathway[69], the fact that they are also produced in the absence of HCN in the starting ice[65] deviates from a Strecker-type synthesis and indicates that they can be formed by radical-radical or radical-molecule reactions[199]. Furthermore the detected β-alanine and diamino acids[65] cannot be made by Strecker synthesis.

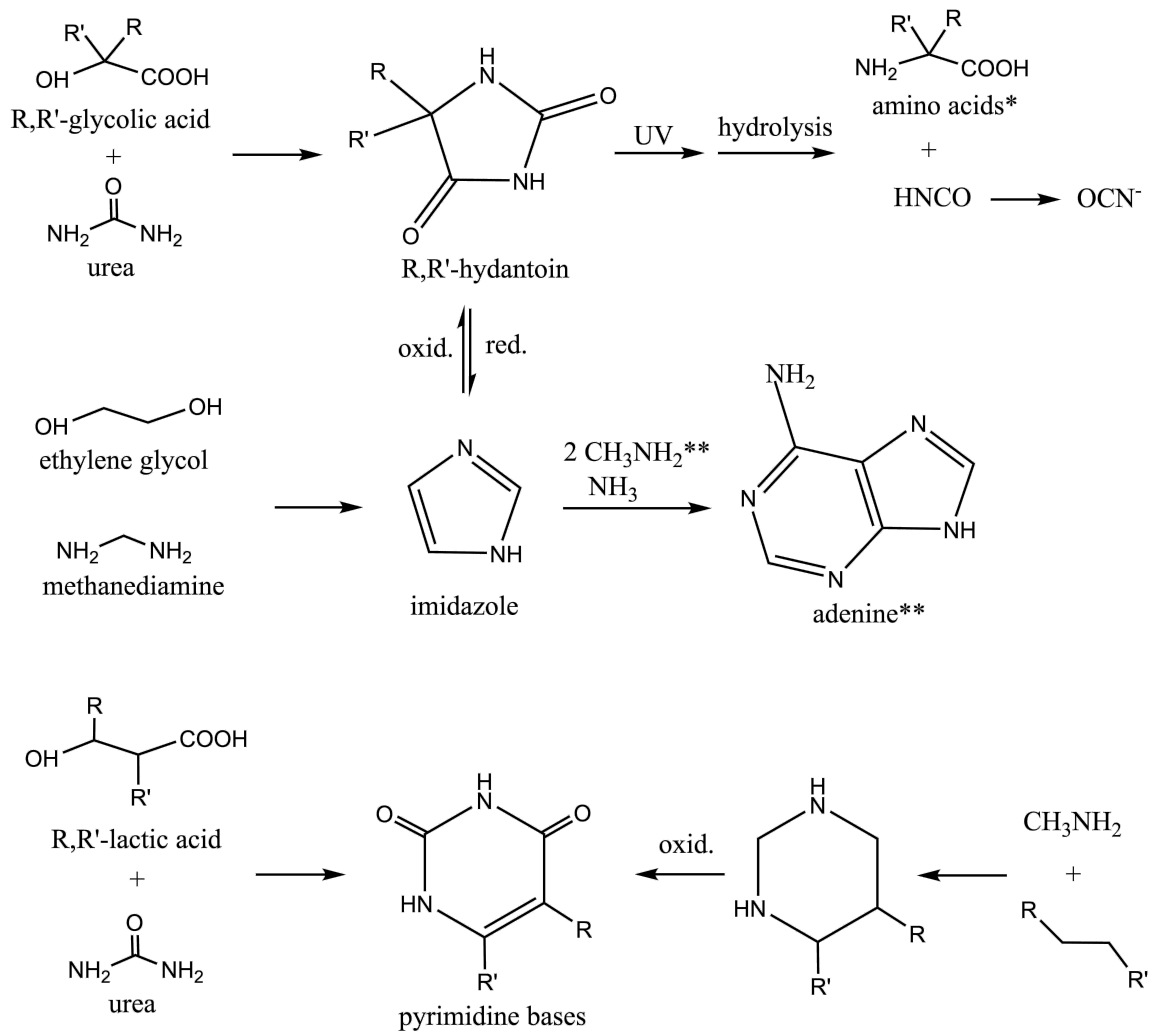

**Fig. 4. Example of the posible formation of N-heterocycles detected in residues.** The proposed reaction pathways need to be confirmed experimentally. It is mediated by other residue components. Upon irradiation or hydrolysis, these N-heterocycles can produce the amino acids reported in the literature, see main text for references.
* Depending on the R,R' groups, any aminoacid could be formed via this process.
** Adenine was chosen as a representative example of purine bases. Changing methylamine by similar reactants, such as HNCO, would produce different purines.

**Nitrogen-heterocycles**
As already mentioned, hexamethylenetetramine (HMT) can be identified in the infrared spectra of residues, and constitutes the most abundant species detected by GC-MS without derivatization[61,62,200,63,74]. HMT-based species have also been detected in residues, these are $C_6H_{11}N_4$–$CH_3$, $C_6H_{11}N_4$–OH, $C_6H_{11}N_4$–$CH_2OH$, $C_6H_{11}N_4$–NH–CHO and $C_6H_{11}N_4$–CHOH–CHO[74,201,202]. Meierhenrich et al. 2005[75] reported the identification of several Nitrogen-heterocycles in the residues using GC-MS with derivatization. Among them, the most abundant one was hexahydro-1,3,5-triazin, also known as trimethylentriamine (TMT), which is a precursor of HMT. Some of these species could act as precursors of biological cofactors[203,75]. The synthesis of TMT was studied[204,177] and procedes by trimerization of methylenimine ($H_2C=NH$), its formation temperature varies between 230 K and room temperature depending on the ice composition[63,177]. Hydantoin (c-$CH_2C(O)NHC(O)NH$) was identified by de Marcellus et al. 2011[76] in residues made from $CH_3OH:NH_3$ ice grown and irradiated at 80 K, but its abundance increased considerably if water was included in the starting ice mixture. Ruf et al. (2019)[78] identified cytosine in residues made under similar conditions. Starting from $H_2O:CO:NH_3:CH_3OH$ ice deposited and irradiated at 10 K, Oba et al. 2019[77] reported the presence of nucleobases in the residues with trace abundances, while imidazole accounted for most of the N-heterocyclic fraction. The strong decrease observed in the abundance of N-heterocycles in residues when hydrolized with 6M HCl at 110 °C, and the enhancement of the amino acid abundances upon hydrolysis, suggests that N-heterocycles act as amino acid precursors. For instance, the synthesis of glycine from hydrolysis of hydantoin is efficient[205]. On the other hand, despite the presence of imidazole in the residue, in line with Miller-Urey-type experiments[206], no amino acids such as histidine that contain imidazole were reported in these residues. A posible explanation for the presence of various residue components is provided in Figure 4. Amines (or amides) perform nucleophilic attacks. Carboxylic acids are very good receptors, since the carbon of the acid group is very susceptible to suffering a nucleophilic attack, giving rise to the corresponding carbonyl derivative. When the acid and amine (amide) skeletons have 5-6 members, cyclization occurs very readily, giving rise to a wide variety of N-heterocycles, such as imidazole groups, hydantoin, pyrimidines, etc.

**Sugars**

As expected, the $C-H_x$ with x=1-3 stretching modes of aliphatics lead to an absorption between 3000-2800 $cm^{-1}$ in the residues made from irradiation of ice mixtures with variable composition. The double-peaked band profile of this band is characteristic of $CH_2$ groups adjacent to OH groups as in glycols, while $CH_3$ groups are in comparison significantly less abundant[207]. This interpretation was later supported with the detection of several sugar molecules in $H_2O:CH_3OH:NH_3$ residues using GC-MS, among them were ribose and various structurally related species. The majority of them included $CH_2OH$ groups along the chain and at the endings, with lower abundances of molecules that contained CHO or COOH functional groups[72]. The formation of these species is linked to the presence of methanol photoprocessing forming the $CH_2OH$ radical. The presence of deoxyribose and deoxysugar derivatives in these residues was more recently reported[73].

**Carboxylic acids**

Carboxylic acids are detected as abundant residue components using GC-MS[60,61]. In order of abundance these are glycolic acid ($HOCH_2COOH$), glyceric acid ($HOCH_2CH(OH)COOH$), 3-hydroxypropionic acid ($HOCH_2CH_2COOH$), and oxamic acid ($NH_2COCOOH$). The relatively high amount of glycolic acid and the presence of glyceric acid in the residues was confirmed, along with other carboxylic acids related to sugars where the $CH_2OH$ group in one end is replaced by a COOH group[72].

**Ammonium salts**

The C=O stretch in carboxylic acids displays a strong IR band near 1700 $cm^{-1}$, but typically this vibration mode in residues is associated to the most prominent band at 1586 $cm^{-1}$, showing evidence for ammonium salts of carboxylic acids ($[NH_4^+][R-COO^-]$)[63]. The formation of these salts occurs spontaneously at 10 K if $NH_3$ is codeposited with a carboxylic acid[63]. The presence of other salts in the ice was also inferred experimentally, an example is $[(OCN^-)(NH_4^+)]$[189,191]. The radiolysis of [(SH$^-$)(NH$_4^+$)] was reported[208], this compound was recently proposed as a component in comet 67P[209].

**Sulfur chains**

As earlier mentioned, the sulfur depletion problem refers to the low abundance of this element in dense interstellar regions, which could be hiding in the dust[210]. S-molecules formed by irradiation of H$_2$S-bearing ice mixtures include species starting with H$_2$S$_2$, formed by HS radicals, with increasing number of S atoms, and S$_2$ to S$_8$ made by their dehydrogenation due to photolysis[153]. Carrascosa et al. 2024[157] shows an efficient formation of primarily H$_2$S$_x$ species with x=2-9 in similar experiments. Shingledecker et al. 2020[211] calculations support a reduction of H$_2$S driving the increase of S-allotropes and specially S$_8$. S$_8$ is an abundant residue component and S$_6$, S$_7$, pentathian (S$_5$CH$_2$), hexathiepan (S$_6$CH$_2$), and c-(S-CH$_2$-NH-CH$_2$-NH-CH$_2$) could also be identified[54].

## Application to astronomical observations of ice and COMs in extraterrestrial samples

In this review, the study of chemical reactions that occur in multicomponent ice mixtures exposed to radiation refers to interstellar and circumstellar icy grains. In the solar nebula, agglomeration of these particles led to comets and asteroids that still preserve to some extent the heritage of COMs made earlier during ice processing, while the composition of larger bodies has been modified more drastically. Nevertheless, the ice layers in planets, moons, or Transneptunian Objects (TNOs) also undergo radiative and thermal processing and can benefit from similar experimental studies[212-214,112,215].

In the infrared, bands are adscribed to functional groups common to different COMs, and they often overlap with the strong absorptions of the simple ice molecules, thus hindering the identification of individual COMs. For instance, it would be difficult to detect glycine in astrophysical ices with some certainty[217]. Recent JWST observations of ice with COMs are currently under study. The presence of a COM in young protostars, CH$_3$OCHO, was classified as ¨secure¨, while that of CH$_3$COOH is ¨likely¨[150].

Compared to interstellar ice, the more detailed chemical characterization of comets allows comparison with ice photoproducts, and they show a high resemblance that unveils the radiation history of precometary ice particles. Calculations suggest that a significant load of organic species in comets and asteroids was delivered to earth during the epoch of heavy bombardment until about 3.9 Gyr ago. The emergence of life forms in our planet could have benefited from these sources of water and prebiotic matter[6,7]. This cycle of organic matter in space is shown in Figure 5.

**Interplanetary dust, meteorites and asteroids**

The amorphous carbon fraction of interplanetary dust particles (IDPs) collected in the stratosphere is often highly enriched in D, $^{15}$N, and $^{17,18}$O, suggesting a formation at very cold temperatures, presumably by processing of sub-micron icy grains[217]. Cluster IDPs formed by agglomeration of these grains experienced heating in the protosolar disk, which lead to conversion of organic species in the ice to amorphous carbon[217]. Organic globules found in meteorites, micrometeorites, and IDPs, could also be the relics of icy grain irradiation, with elevated D/H and $^{15}$N/$^{14}$N ratios and wall thicknesses that roughly correspond to the UV-photon penetration in the ice[218,219]. Among meteorites, carbonaceous chondrites like Murchison contain amino acids and other organic species with high D/H ratios, an indication of their formation at the low temperatures typical of icy grains[220]. This meteorite also contains diamino acids[195], HMT and HMT-derivatives[221] that coincide remarkably with those made by ice irradiation[65,74], see Sect. Amino acids).

Recently, the Hayabusa2 mission returned to Earth samples from carbonaceous asteroid Ryugu. This minor body is rich in organic molecules that suggest thermal and/or aqueous alteration, comprising amino acids, uracil, carboxylic acids (HCOOH and $CH_3COOH$), amines (in order of abundance $CH_3NH_2$, $C_2H_5NH_2$, $(CH_3)_2CHNH_2$, and $C_3H_7NH_2$ likely present as salts) and N-heterocycles (pyridine, piperidine, pyrimidine, imidazole, and pyrrole rings with various amounts of alkylation)[222,223]. In a recent paper[224], IR spectra of nanoglobule-like inclusions in Ryugu samples are compared to the residues made by irradiation of interstellar ice analogs[63]. These authors conclude that the composition of these nanoglobules, enriched in C=O and $CH_x$, are compatible with a formation by ion or UV irradiation of ice mixtures. The presence of $S_8$ in residues made from $H_2S$-bearing ice irradiation is common to the Orgueil meteorite and asteroid Ryugu samples[225]. Meanwhile, a preliminary analysis of returned Asteroid (101955) Bennu samples has been published [226].

**Comets**
Compared to asteroids, comets have experienced a lower degree of aqueous and thermal alteration, being therefore a better match of organic species made by photoprocessing of interstellar ice. The COSAC instrument on board the Philae lander of Rosetta performed the first analysis of a cometary nucleus, collecting a mass spectrum that provides the tentative identification of 16 species[227,228], these are $H_2O$, $CH_4$, HCN, CO, $CH_3NH_2$, $CH_3CN$, HNCO, $CH_3CHO$, $HCONH_2$, $C_2H_5NH_2$, $CH_3NCO$, $CH_3COCH_3$, $C_2H_5CHO$, $CH_3CONH_2$, $HOCH_2CHO$, and $(CH_2OH)_2$. From this list, $CH_3NCO$ and $CH_3CN$ are related to $OCN^-$ and HCN, frequently found in ice irradiation experiments. It may not be casual that most of these species are among the photoproducts of interstellar ice analogs and, furthermore, they can be synthesized by reactions of the ice radicals: OH, NH, $NH_2$, $CH_3$, HCO, and $CH_2OH$. Unfortunately, the non-nominal landing of Philae did not allow a more sophisticated analysis involving GC-MS[229]. The ROSINA instrument on board the Rosetta orbiter confirmed two molecules inferred from the COSAC data, methylamine ($CH_3NH_2$) and ethylamine ($CH_3CH_2NH_2$)[230]; as mentioned above these are also the most abundant amines in the Ryugu asteroid. Among other findings, they proposed that glycine was present in the comet dust (in line with the earlier glycine detection in cometary samples returned by Stardust[231]), sulfur molecules $S_2$ to $S_4$ or larger species containing such fragments, and ammonium ($NH_4^+$) salts with $CN^-$, $OCN^-$, $Cl^-$, $HCOO^-$, $CH_3COO^-$, and $SH^{-}$[151,232,209]. This converges very nicely with the presence of these molecules in organic residues made from ice photoprocessing. S allotropes from $S_2$ to $S_8$ are readily made by irradiation of ice containing $H_2S$ and their presence in the comet of Rosetta was predicted[54]; unfortunately the detection of $S_8$ during the Rosetta mission was not technically possible, but the detected $S_2^+$ to $S_4^+$ cations could be originated either from $S_2$ to $S_4$ molecules, or fragments of larger S allotropes[153]. We mentioned that $S_8$ has been recently detected in the Orgueil meteorite and asteroid Ryugu samples[225]. Ammonium salts could also be present in interstellar ice[190], we showed that they are an abundant residue component[63]. Furthermore, the anions associated to these salts in the Rosetta detections match those previously identified in the ice irradiation experiments: the $OCN^-$ formed in the ice only desorbs near room temperature when it is part of the ammonium salt[233], carboxylic acids that participate in the cometary salts ($HCOO^-$ and $CH_3COO^-$) are common ice photoproducts, and the more refractory ones like the glycolic acid anion (HOCH$_2$COOHO$^-$) are among the most abundant residue components, see Sect. Carboxylic acids. Finally, the high abundance of H$_2$S in comet 67P favors its participation in the ammonium salts.

Spectroscopy in the infrared is a common tool to observe solids in space. The double-peaked IR absorption band profile at 3000-2800 cm$^{-1}$ (known as the 3.4 μm feature), caused by CH stretching bonds in residues made from realistic ice mixtures, is shared with glycols (OH(CH$_2$)$_n$OH), and was therefore attributed to CH$_2$ stretchings altered by adjacent –OH groups in the residues[207]. This assigment was reinforced with the detection of sugar molecules in the residues, among them ribose and other related species[72]. The 3.4 μm feature could serve as a tracer of organics made by ice irradiation in space, although it overlaps with the strong water absorption in ice. This band was used to fit the IR spectra of comet 67P nucleus by the VIRTIS/Rosetta spectrometer[234], but the cometary feature was shallow and did not allow a robust identification.

**Other types of radiation for ice processing in space**
Most of the discussion presented in this review refers to UV radiation of ice in space. Other sources of ice processing, X-ray photons and ionizing particles, tend to produce similar products in the ice[23-28,81]. In general, it is not obvious to associate the chemical abundances in an astrophysical environment to the local radiation source. One exception could be protoplanetary disks, where X-rays are expected to play a pivotal role in photochemistry and photodesorption. X-ray processing of a realistic ice sample, composed of a layer of H$_2$O:CH$_4$:NH$_3$ covered by a layer of CO:CH$_3$OH, provided a good match of the observed simple species in these disks (CO, HCO, H$_2$CO, but no CH$_3$OH), while the non-detection of COMs is explained by the negligible desorption of these species made in the ice bulk[115].

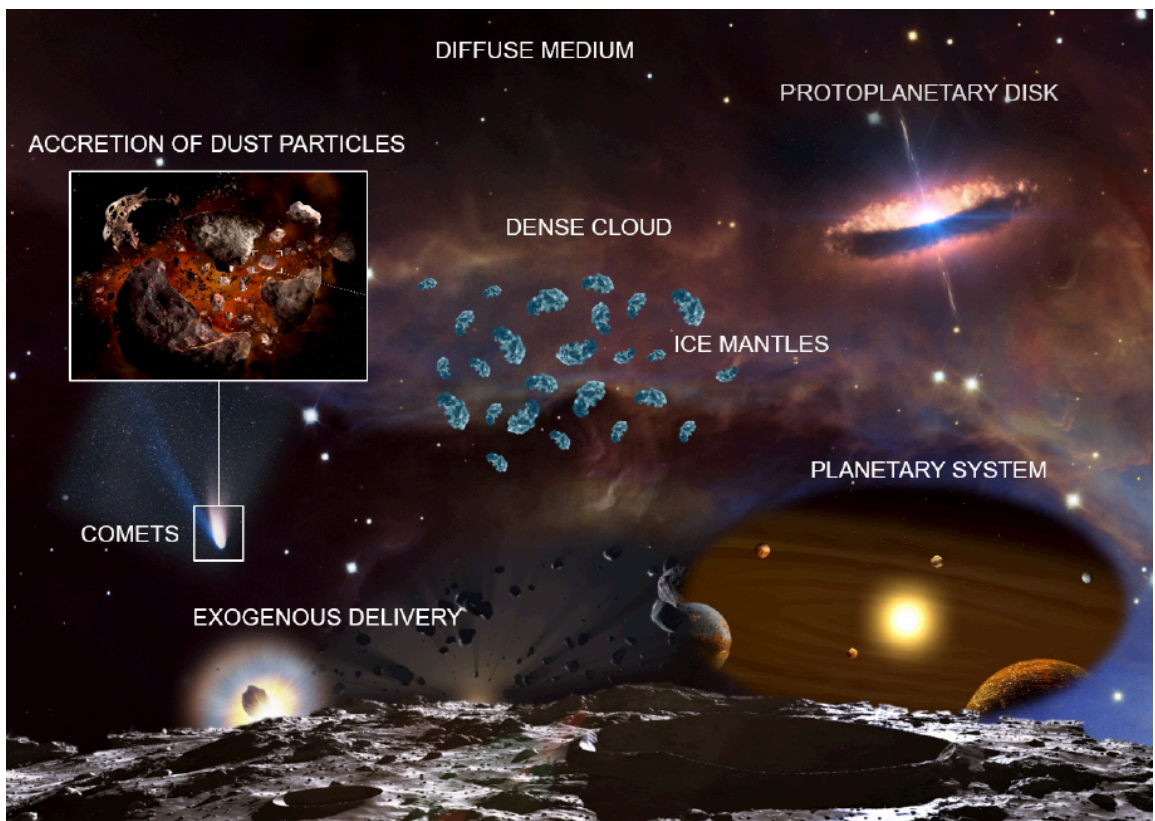

**Fig. 5. Organic species of prebiotic interest made in space.** This figure shows the cycle of matter in the interstellar medium. Diffuse clouds can collapse to create a dense cloud, where ice mantles accrete onto dust grains and molecular chemistry takes place forming COMs. Within this cloud, protoplanetary disks are formed, here icy dust agglomeration into cometesimals allows the preservation of COMs in comets and other primitive bodies. Upon impact of these objects, water and COMs were delivered to the early earth. Eventually, the death of the star will return this matter to the interstellar medium, starting a new cycle. Credit: Composite includes ESO website images.

## Conclusions and outlook

The journey of relatively abundant C, H, O, N, S atoms in space is a long and complex one. After their stellar synthesis, these atoms permeate the interstellar medium and react in the gas phase to form some of the observed molecules, while other species can only be produced on the surface of dust particles, where they also accrete to build ice mantles. In the interior of dense clouds the formed ice is exposed to a low (but long-lasting) UV-radiation field and to direct cosmic-ray impact. In protoplanetary disks, the very young star emits mainly X-rays and later UV photons. The large variety of processes promoted by radiation, and by occasional heating, of the ice leads to a bountiful of complex organic species. We discussed that many of them are of astrobiological interest. Table 2 provides a list of molecules with 6 to 13 atoms detected in various space environments or synthesized in laboratory simulations. Since they contain at least one C atom, they are all COMs. The cometesimals, formed by accretion of tiny icy dust grains in circumstellar disks, preserve this organic heritage and continue to be exposed to ion processing. It is thus no longer surprising that comets contain organic compounds similar to those synthesised in the laboratory by irradiation of interstellar ice analogs. The study of carbon-rich meteorites and asteroids suggests that these bodies also contain primordial organic matter that underwent further processing, and sometimes also aqueous alteration. As earlier mentioned, the delivery of these materials to the early earth via cometary and asteroidal impacts likely contributed to the emergence of the first life forms. While these bodies also carry highly destructive effects, microscopic dust aggregates present in the protoplanetary disk likely provided a gentler deposition of organic material, as IDPs still do today[7]. Because this evolutionary sequence of matter in space operates in the Milky Way and in other galaxies, exoplanets should be shreaded by similar organic molecules, fostering life if proper conditions are met in the host world. This scenario was envisioned by pioneering minds before the launch of the first space missions. What is new is the confirmation of these ideas based on our study of pristine bodies, theoretical and laboratory simulations of space environments that conform what we know as Astrochemistry. Future missions to comets and asteroids, including sample return missions, spectroscopic observations of ice environments, and more realistic theoretical/laboratory simulations will continue to unveil the complex chemistry of ices in space, and tell us a bit more about our own origins.

## Acknowledgements

This research has been funded by projects PID2020-118974GBC21 and PID2023-151513NB-C21 by the Spanish Ministry of Science and Innovation. R.M.-D. has received funding from "la Caixa" Foundation, under agreement LCF/BQ/PI22/11910030.I. We acknowledge advice from Dr. Y.-J. Chen and Dr. S. Cazaux, and constructive comments from the reviewers that contributed to improve this review.

## Author contributions

G. M. M. C. wrote the main text of this manuscript and researched data for the article. H. C. made the figures. H. C. and R. M.-D. contributed to writing, discussion of content, and reviewing the manuscript before submission.

## Competing interests statement

The authors declare no competing interests.

## Graphical abstract

Please provide a suggestion for the graphical abstract as a separate figure file. Often, the graphical abstract can be assembled from several parts of images already used but note that the space is tight; the image must fit in a space that is 9 cm wide by 4 cm high. Our art team will prepare the final figure based on your suggestion. Please avoid using third party images.

## ToC blurb

*This Review highlights the synthesis of species made by ice irradiation in space. Among them are organics of astrobiological interest that were delivered to the early earth via comets and asteroids.*

Table 2 | **Molecules detected toward different environments in space or produced in laboratory simulations of ice photoprocessing with 6 to 13 atoms and at least one C (COMs).** This table is an updated version of a previous one[81] based on a recent publication[235].

| 6 | | 7 | | 8 | | 9 | | 10 | | 11 | | 12 | | 13 | |
|---|---|---|---|---|---|---|---|---|---|---|---|---|---|---|---|
| $CH_3OH^{(*)}$ | •••• | $CH_3CHO$ | •••• | $HCOOCH_3$ | • • • | $CH_3OCH_3$ | • • • | $(CH_3)_2CO$ | • | $HC_9N$ | • • | $C_6H_6$ | •• | $C_6H_5CN$ | • |
| $CH_3CN^{(*)}$ | ••••• | $CH_3CCH$ | •• •• | $CH_3C_3N$ | • | $CH_3CH_2OH$ | • | $(CH_2OH)_2$ | • | $CH_3C_6H$ | • | $n-C_3H_7CN$ | • | $HC_{11}N$ | • |
| $CH_3CO^+$ | • | $C_2H_5OH$ | • | $C_2H_6$ | • | $CH_3CH_2CN$ | • | $HOCH_2CONH_2$ | • | $CH_3CH_2OCHO$ | • | $i-C_3H_7CN$ | • | $HOCH_2CH(OH)CO_2H$ | ♦ |
| $NH_2CHO$ | •••• | $CH_3NH_2$ | • | $C_3H$ | • • • | $HC_7N$ | • • • | $CH_3CHO$ | • | $CH_3COOCH_3$ | • | $CH_3CH(OH)CO_2H$ | ♦ | $c-C_4H_5N_3O$ | ♦ |
| $NH_2CO_2^-$ | | $CH_2CHCN$ | • • • | $CH_3COOH$ | • | $CH_3C_4H$ | • • • | $CH_3C_5N$ | • | $C_5H_8$ | | $HOCH_2CH_2CO_2H$ | ♦ | | |
| $HCOHCO$ | • | $HC_5N$ | •••• | $H_2C_6$ | • | $C_3O_2$ | • | $CH_3CHCH_2O$ | • | $NH_2COO^-NH_4^+$ | | $NH_2CONHCONH_2$ | ♦ | | |
| $CH_2NCO$ | | $C_3O_2$ | • | $CH_2OHCHO$ | • | $C_8H$ | • • | $CH_3OCH_2OH$ | • | $HOCH_2CH_2NH_2$ | ♦ | $1-C_3H_5CN$ | • | | |
| $CH_2CNH$ | | $C_2H_5$ | • | $HC_6H$ | •• | $C_8H^-$ | • • | $NH_2CH_2COOH$ | | $CH_3COCH_2OH$ | • | $2-C_3H_5CN$ | • | | |
| $H_2CO_3$ | | $C_6H$ | •• • | $CH_2CHCHO$ | • | $CH_2CONH_2$ | • | $NH_3^+CH_2COO^-$ | | $C_3H_6$ | • | | | | |
| $CH_3SH$ | • | $C_6H^-$ | • • | $CH_2CCHCN$ | • | $CH_2CHCH_3$ | • | $NH_2COCONH_2$ | ♦ | $c-C_3H_4N_2O_2$ | ♦ | | | | |
| $C_2H_4$ | • | $C_6O$ | | $CH_3CHCH_2$ | | $NH_2CH_2CN$ | • | $CH_3CH_2SH$ | • | $NH_2CONHCHO$ | | | | | |
| $C_4O_2$ | ♦ | $CH_2CHOH$ | | $CH_3CHNH$ | • | $HC_7O$ | • | $NHC(NH_2)NHCN$ | ♦ | | | | | | |
| $C_3O$ | • | $c-C_2H_4O$ | • | $CH_3SiH_3$ | • | $C_2H_5NO$ | ♦ | $C_{10}$ | ♦ | | | | | | |
| $C_5H$ | • | $CH_3NCO$ | • | $C_9$ | | | | | | | | | | | |
| $CH_2NC$ | •• | $HC_5O$ | • | $HCOCONH_2$ | | $C_3H_6$ | | | | | | | | | |
| $HC_2CHO$ | • • | $CH_3NO_2$ | • | $CH_3CH_2NCO$ | • | $HOCH_2CO_2H$ | ♦ | | | | | | | | |
| $C_5S$ | • | $CH_3OCH_3$ | | $NH_2CONH_2$ | • | $NH_2COCO_2H$ | ♦ | | | | | | | | |
| $HC_3NH^+$ | • | $NH_2COOH$ | | $C_7O$ | | $CH_3NHCHO$ | • | | | | | | | | |
| $C_5N$ | • | $NH_4^+CN^-$ | | $C_8$ | | $H_2C_7HC_2H$ | • | | | | | | | | |
| $C_5N^-$ | • | $C_2H_6O$ | ♦ | $C_2H_4O_2$ | ♦ | $HC_5HCHCN$ | • | | | | | | | | |
| $C_6$ | | $HOCH_2OH$ | ♦ | $NH_2CH_2OH$ | ♦ | $H_2C_2HC_3N$ | • | | | | | | | | |
| $HC_4H$ | •• | $C_7$ | • | $S_2CH_2$ | • | $c-C_3H_4N_2$ | ♦ | | | | | | | | |
| $HC_4N$ | • | $HCNHNH_2$ | ♦ | $C_4H_4$ | ♦ | | | | | | | | | | |
| $c-H_2C_3O$ | ♦ | $HOCH_2CN$ | • | $HCCCH_2CN$ | • | | | | | | | | | | |
| $CH_2CNH$ | • | $HC_4NC$ | • | $CH_2CHCCH$ | • | | | | | | | | | | |
| $CH_2CCH$ | • | $c-C_5HCCH$ | • | | | | | | | | | | | | |
| $HNCHCN$ | • | | | | | | | | | | | | | | |

$^{(*)}$ $CH_3OH$ and $CH_3CN$ have been observed in disks
• hot and/or shocked regions • PDRs • dark clouds • diffuse clouds • envelopes of evolved star • external galaxies
♦ Experiment IR ♦ Experiment MS ♦ Experiment GC-MS/HPLC